\documentstyle[12pt,epsf]{article}
\begin{document}
\renewcommand{\textfraction}{0}
\renewcommand{\topfraction}{1}
\title{ 
\vskip -0.3cm
\makebox[12cm]{}{\small WUB 96-43}\\
\vspace{0.6cm}
HEAVY QUARK PHYSICS ON THE LATTICE\footnote{Invited talk given at 
HEAVY QUARK PHYSICS AT FIXED TARGET, St. Goar, Oct. 3rd, 1996.}}
\author{
Stephan G\"usken \\
{\em Physics Department, University of Wuppertal}\\
{\em D-42097 Wuppertal, Germany} \\
{\em E-mail: guesken@theorie.physik.uni-wuppertal.de} 
}
\date{}
\maketitle
\baselineskip=14.5pt
\begin{abstract}
We illustrate the current status of heavy quark physics on the lattice.
Special emphasis is paid to the question of systematic uncertainties
and to the connection of lattice computations to continuum physics.
Latest results are presented and discussed with respect to the progress
in methods, statistical accuracy and reliability. 
\end{abstract}
\baselineskip=17pt
\section{Introduction}
The role of lattice QCD in the field of heavy quark physics is
twofold.

First of all it provides experimental and
phenomenological physicists with reliable predictions of those
non-perturbative QCD parts, which are in nature almost inevitably
connected to most of the important electroweak processes.
For example, the Cabbibo-Kobayashi-Maskawa  matrix elements for
the weak decay of the $b$ quark cannot be extracted from 
experimentally measured B meson decays without distinct knowledge
about QCD bound state properties of heavy light systems, i.e. the
form factors.    

Secondly, in the heavy quark regime lattice QCD can 
successfully  describe "pure QCD" quantities like
quarkonia splittings and the heavy quark potential,
which, in turn, can be used to determine
the strong running coupling constant $\alpha_s$. 

In order to judge on the quality of results from lattice
QCD it is necessary to understand roughly how the lattice method
works in practice, what are its advantages and shortcomings,
where systematic
uncertainties originate from and how these uncertainties can be
controlled.

In principle, the lattice procedure consists
in solving the QCD path integral on a (Euclidean) space time
lattice.  
The connection with continuum physics is made in the limit, where  
(a) the lattice constant $a$ goes to zero, and
(b) the lattice volume $V$ goes to infinity. If it were possible
to achieve this limit unambiguously, QCD would be solved.

In practice, however, the path integral can be evaluated only
for  finite values of $a$ and $V$.   
This is done numerically with the help of
Monte Carlo methods,
which introduce a statistical uncertainty on the results.
In order to achieve the continuum limit, extrapolations
in $a$ and $V$ are necessary.
On top of this, most of the lattice calculations in the past
have been performed in the so called quenched approximation, where
(roughly speaking) internal fermion loops are neglected.

Progress in lattice QCD therefore has to be measured in terms of
\begin{itemize}
\item[(a)] Statistical significance
\item[(b)] Reliability of the extrapolation 
$a \rightarrow 0$, $V\rightarrow \infty$
\item[(c)] Ability to include  internal fermion loops. 
\end{itemize} 

Especially in the case of heavy-light systems, 
item (b) confronts lattice QCD with a serious problem.
The inverse of the lattice constant $a^{-1}$ can be viewed 
as a (gauge invariant) cutoff to all lattice observables.    
With todays computer facilities, cutoffs up to 
$a^{-1} \simeq 3.5 \mbox{GeV}$ can be achieved for suitably large
lattice volumes. If one would try to calculate B-meson properties
, with $M_B \simeq 5 GeV$, directly on such a lattice, severe
finite cutoff effects would prevent a reliable extrapolation
$a\rightarrow 0$.

An obvious way to alleviate this problem is of course to extrapolate
lattice data in mass from far below the cutoff into the region of
heavy masses, e.g. to $M_B$.
However, as the functional dependence of the lattice data on
the heavy quark
mass is in general not exactly known, this procedure introduces 
additional systematic errors, which need to be controlled. 

In order to overcome this unsatisfactory situation, several methods 
have been developed and improved over the recent years.
These methods can be classified into three different groups:
a priori methods, phenomenological methods, and effective methods.

A priori methods trie to improve on the discretization
of the QCD action in such a way, that cutoff effects are reduced.
Those "improved actions"  converge faster to the correct
continuum form, leaving the QCD physics unchanged.
The most prominent example\footnote{We comment here only 
on the fermionic part of the QCD action. For improvements on
the gluonic part see 
refs.\cite{improved_gluon_luescher,improved_gluon_lepage,SHW_action_tadpole}.}
 in this group is the Sheikholeslami-Wohlert
action\cite{SHW_action}, which is often called clover action.
In contrast to the standard Wilson action\cite{W_action}, finite
cutoff effects proportional to $a$ are absent, at least
on the classical level.
It turns out, however, that quantum fluctuations can re-introduce
$O(a)$ cutoff effects. This problem is currently tackled by
a proper adjustment of the coefficient of the (additional) clover
term. Lattice actions which include this adjustment are traded under the
names "tadpole improved clover action"\cite{SHW_action_tadpole} or
"non-perturbatively improved clover action"\cite{SHW_action_nonpert}.\\
In principle it should be possible to formulate a "perfect"
lattice action, 
which is free of any cutoff effect. The construction of such an action
is currently 
under consideration\cite{Wiese_perfect}.

The basic idea of phenomenological methods is, to reduce
finite cutoff effects by proper adjustment of lattice observables.
This is  achieved by a change in the normalization of the lattice
quark propagator (LMK I)\cite{LMK_scheme} and by a redefinition of the
particle masses (LMK II)\cite{LMK_scheme_improved}. As the modifications
are not implemented on a fundamental level, it is not clear whether
this method  leads to a general improvement of all lattice
observables.

Effective methods have been designed for heavy quarks
on the lattice. The idea is to remove the largest scale, i.e.
the heavy quark mass, from the lattice action. If all remaining
scales are small compared to the lattice cutoff, finite $a$ effects
should be reduced substantially. 
Lattice implementations of effective methods have been developed for
(a) the static approximation\cite{Static_app},
i.e. the zeroth order of heavy quark effective theory, and (b) 
the non-relativistic QCD\cite{NRQCD} (NRQCD). The latter approach
starts from a non-relativistic approximation of QCD and includes,
similar to the well known Fouldy-Wouthusen transformation,
successively relativistic corrections in form of a 
$1/M_Q$ expansion. Clearly,  the range of
validity of such methods is limited to the region of (very)
heavy quarks.

In view of the variety of all these methods, whose merits
in some cases have not been fully proven yet, one could argue that 
lattice QCD has lost its status of being an ab initio method.
In the following we will try to convince you of the opposite.

The backbone of large scale lattice calculations is always
a powerful computer. The strong increase
in sustained computer speed over the recent years
allowed for a substantial improvement on statistical significance of
lattice data as well as for a variation of  the lattice constant and the
lattice volume over a much larger range.
With such a tool in hands, one can indeed check on the benefit of
the different methods and calculate a reliable estimate  of the
size of systematic uncertainties. For example, one can measure
the size of finite cutoff effects by performing a series of
lattice simulations with different lattice constants $a$.    
In that view, the computer can be compared with the accelerator
and the detector of a large experiment: the quality of the
device has an decisive influence on the quality of the results. 

In the following the actual status of the "lattice experiments"
within the field
of heavy quark physics is reviewed in form of
selected examples. We will discuss the decay constant of the $B$ meson,
$f_B$,
the semileptonic decays of $B$ mesons, and the determination of $\alpha_s$
from quarkonium splittings and the heavy quark potential.

\section{Status of $f_B$} 
\subsection{Strategy}
The decay constant of the $B$ meson, $f_B$, is defined in
the space time continuum by
\begin{equation}
\langle 0| A_0 |B \rangle^{cont} \equiv m_B f_B \quad ,
\label{ZA_def}
\end{equation}
where $A_0$ denotes the zeroth component of the axial current and
$m_B$ is the $B$ meson mass. Thus, $f_B$ can be determined once the
matrix element on the left hand side of eq.(\ref{ZA_def}) is known.

On the lattice, one calculates $\langle 0| A_0 |B \rangle^{latt}$,
which is related to its pendant in the continuum by
\begin{equation}
\langle 0| A_0 |B \rangle^{cont} = Z_A \langle 0| A_0 |B \rangle^{latt}\;.
\label{ZA_latt_cont}
\end{equation} 
The renormalization constant $Z_A$ accounts for the non-conservation
of the axial current. Both,  $\langle 0| A_0 |B \rangle^{latt}$
and $Z_A$ depend on the lattice spacing $a$, and much work
has been devoted to
the problem of choosing $Z_A$ such that it just cancels the finite cutoff
effects of the above 
product\cite{LMK_renorm,Martinelli_renorm}.   
As a satisfactory solution to this problem has not been found yet, an
extrapolation $a \rightarrow 0$ is still necessary.
\begin{figure}[htb]
\epsfxsize=12.0cm
\centerline{\epsfbox{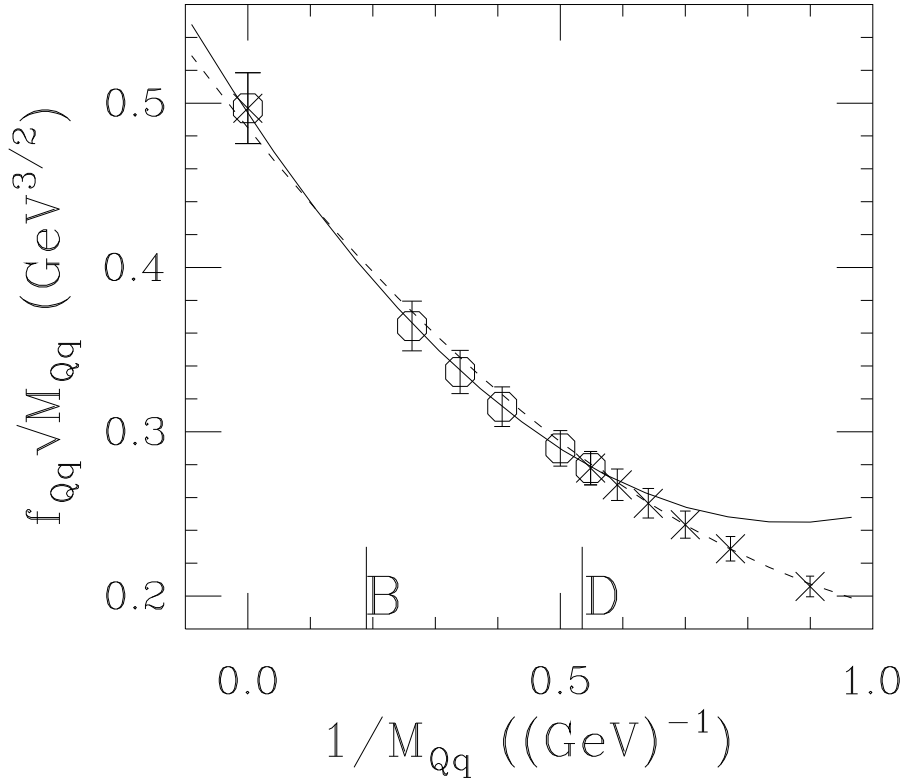}}
\caption[a]{\label{fig_MILC_fb_mps}
\it MILC collaboration\cite{MILC_fb}: $f_{Qq}\sqrt{M_{Qq}}$
vs. $1/M_{Qq}$ at a finite value of the cutoff ($a^-1 \simeq 3.2\mbox{GeV}$).
The solid line is a quadratic fit to the octagons
( "heavier heavies" + static); the dashed line
is a quadratic fit to the crosses ("lighter heavies" +static). The
difference of the two curves in the heavy quark region illustrates
the quality of the interpolation.}
\end{figure}

Due to limitations in computer time -- and computer memory size --
the $B$ meson mass is beyond currently attainable lattice cutoffs.
Therefore, one calculates $f_{PS}$ with meson masses below the cutoff
and in the static limit. An $1/m_{PS}$ expansion then interpolates
between these results, yielding $f_B$ at a given value of the lattice
cutoff.  The quality of such an interpolation
is demonstrated in fig.\ref{fig_MILC_fb_mps}.

This procedure has to be repeated for
several cutoff values,
and finally the continuum extrapolation $a \rightarrow 0$ has to be
performed\footnote{A continuum extrapolation includes in general
also the limit $V \rightarrow \infty$. As the finite volume effects are
not crucial for heavy quarks,
we will not discuss this point in detail.}.      

As an alternative to this standard procedure, one can in principle use
the NRQCD effective method. Unfortunately, NRQCD is a
non-renormalizable theory and one cannot send $a\rightarrow 0$.
Therefore, one needs good control over discretization errors of the
NRQCD action. In the end it is necessary to verify, that the results
from the standard method and from NRQCD are consistent.   
  
\subsection{Results}
Three years ago, the first calculation, which includes the
$a\rightarrow 0$ extrapolation\footnote{Results at fixed cutoff have
been published by \cite{fb_BLS,fb_UKQCD,fb_ELC}.}, has been
performed by the PSI-WUPP
collaboration\cite{fb_PSI_WUPP}. In this simulation, the cutoff and
the lattice volume 
were varied in the ranges 
$ 1.5\mbox{GeV} \leq a^{-1} \leq 3.0\mbox{GeV}$ and
$ 1.0\mbox{fm} \leq V \leq 1.5\mbox{fm}$ respectively. The continuum
extrapolation yielded $f_B = 180(50)$MeV.

This year, the lattice determination of $f_B$ has been improved by
large scale simulations of the JLQCD group\cite{JLQCD_fb} and the
MILC collaboration\cite{MILC_fb}. Preliminary results of a 
NRQCD calculation of $f_B$ have been published by the SGO
collaboration\cite{SGO_fb}.

\begin{figure}[htb]
\epsfxsize=14.0cm
\centerline{\epsfbox{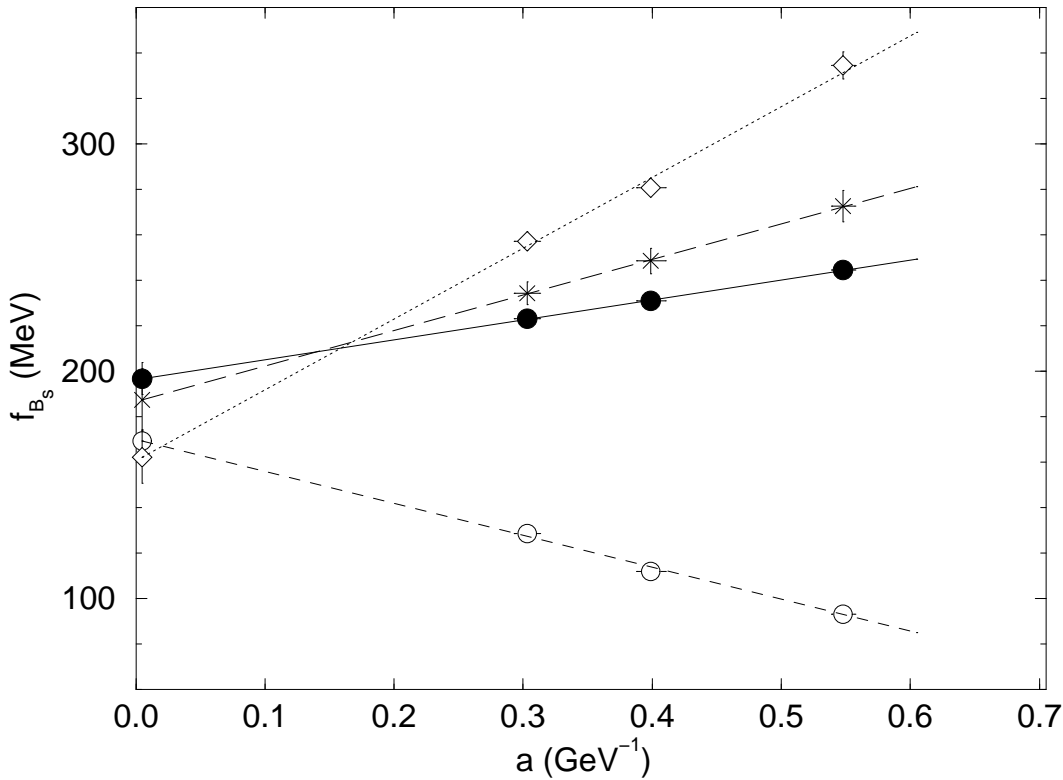}}
\caption[a]{\label{fig_JLQCD_fb_a}
\it JLQCD collaboration\cite{JLQCD_fb}: $f_{B_s}$ as a
function of the lattice spacing $a$. Different symbols refer to
different choices of $Z_A$: naive (open circles), KLM (open diamonds),
KLM improved (crosses), mixed KLM (filled circles).}
\end{figure}
JLQCD has pushed forward in the size of cutoff and volume, 
namely $ 1.8\mbox{GeV} \leq a^{-1} \leq 3.4\mbox{GeV}$ and
$ 1.7\mbox{fm} \leq V \leq 2.0\mbox{fm}$.
An important ingredient of their analysis is a careful study
of the influence of different choices of $Z_A$ on the cutoff
dependence of $f_B$. The result is displayed in 
fig.\ref{fig_JLQCD_fb_a}. It turns out that, although none of the
choices of $Z_A$ removes the cutoff dependence completely, the
continuum values agree within errors. The remaining
systematic uncertainty due to $Z_A$ can be read off the 
spread of results at $a = 0$.
The final JLQCD results read
$$
f_B = 179(11)^{+2}_{-31}\mbox{MeV} \qquad
f_{B_s} = 197(7)^{+0}_{-35}\mbox{MeV} \; .
$$
The first error accounts for statistical, the second for
systematic uncertainties. 
It is very encouraging to see the PSI-WUPP result being consolidated
by the -- more accurate -- JLQCD calculation.
\begin{figure}[htb]
\epsfxsize=14.0cm
\centerline{\epsfbox{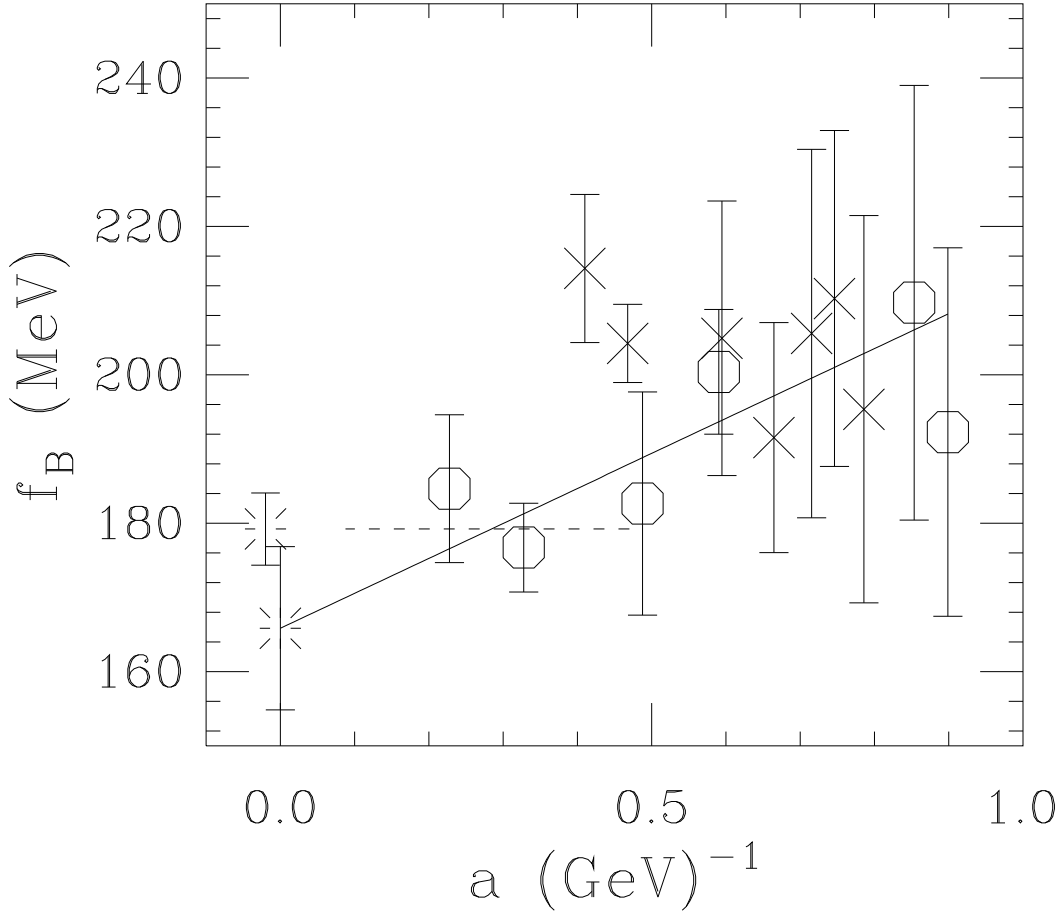}}
\caption[a]{\label{fig_MILC_fb_a}
\it MILC collaboration\cite{MILC_fb}: $f_{B_s}$ as a
function of the lattice spacing $a$. Octagons refer to quenched
data; crosses to full QCD, $n_f=2$. The solid line is a fit to
all quenched points (conf. level = 0.66); the dashed line is a
constant fit to the three quenched points with $a < 0.5\mbox{GeV}$
(conf. level = 0.76). The extrapolated values at $a=0$ are indicated
as bursts.}
\end{figure}

The MILC collaboration has extended the cutoff even further.
Their largest value is $a^{-1} \simeq 4.5\mbox{GeV}$.
Fig.\ref{fig_MILC_fb_a} shows the final data as well as the
extrapolation to $a=0$. For the first time in
lattice QCD, MILC has been able to include full QCD simulations 
-- with $n_f =2$ dynamical flavors -- into the determination of
$f_B$. The corresponding data is represented by crosses in 
fig.\ref{fig_MILC_fb_a}. For small $a$ the full QCD data seem to 
enhance the value of $f_B$. However, the statistical uncertainty 
is too large to draw a firm conclusion. Instead, MILC quotes
the quenched result and includes the effect of dynamical fermions
as a systematic uncertainty
$$
f_B = 166(11)(28)(14)\mbox{MeV} \qquad
f_{B_s} = 181(10)(36)(18){MeV} \; .
$$    
The first error includes statistical errors and systematic effects of
changing fitting ranges, the second other systematic errors within
the quenched approximation, and the third accounts for quenching 
effects. 
Within errors, the MILC result is well compatible with PSI-WUPP and
JLQCD.

The SGO collaboration has analyzed $n_f =2$ full QCD configurations,
using NRQCD for the heavy quark. They quote a (preliminary) value of
$f_B \simeq 180\mbox{MeV}$.

In summary, the "old" value $f_B = 180(50)\mbox{MeV}$ has been
consolidated by more advanced lattice simulations.
However, much work remains to be done in order to reduce 
statistical and systematic uncertainties. A major goal of the next
years will be the inclusion of dynamical fermion loops. 

\section{Semileptonic Decays of $B$ Mesons}

Semileptonic decay amplitudes can be described as a product of a 
(perturbatively accessible) weak interaction term and perturbatively
not accessible QCD part. The latter is commonly parameterized
by a set of form factors\cite{formfactor_def}
\begin{eqnarray}
\langle \mbox{PS'}|J_{\mu} |\mbox{PS} \rangle &=& 
 \mbox{Expr}\left\{ f^+(q^2),f^0(q^2)\right\} \label{ffs_def} \\
\langle \mbox{V}|J_{\mu} |\mbox{PS} \rangle &=& 
 \mbox{Expr}\left\{ V(q^2),A_1(q^2),A_2(q^2),A(q^2)\right\}\; . \nonumber
\end{eqnarray}    
PS, PS' and V are momentum dependent pseudo-scalar and vector
meson states, and $J_{\mu}$
is the (axial-) vector current.  
Clearly, the contribution from lattice QCD is a determination 
of the form factors
by a calculation of the left hand side of eq.(\ref{ffs_def}).
Compared to the calculation of $f_B$ this is not an easy task, as
both, the mass dependence and  the $q^2$ behavior has to be
determined. 
Due to limited  statistical accuracy, it is 
not yet possible to extract the functional behaviors
unambiguously from the data. Thus, systematic uncertainties due
to the choice of inter- and extrapolation methods have to be 
estimated and finally included into the results. 

\subsection{$B \rightarrow \pi,\rho$}
Considerable progress with respect to the reliability
of inter- and extrapolations has been achieved within
the last two years.  "Old" methods\cite{ELC_ffs,APE_ffs},
simply relied on the validity of the pole dominance hypothesis
and on HQET. As the latter is justified  only in the region
$q^2 \simeq q^2_{\mbox{max}}$ one had to operate at very large
values of $q^2$.  

As current lattice investigations substantially improved on
the statistical quality of data, it has become possible to work at
more moderate $q^2$ values and  to check on the influence
of the various assumptions.

The Wuppertal group\cite{ffs_Wuppertal} has performed
its analysis at small values of $q^2$ and for meson masses
up to $M_D$.
In this region the form factors  
can be  determined quite reliably. The extrapolation in mass
\begin{figure}[htb]
\noindent\parbox{15cm}{
\parbox{7.5cm}{\epsfxsize=7.5cm\epsfbox{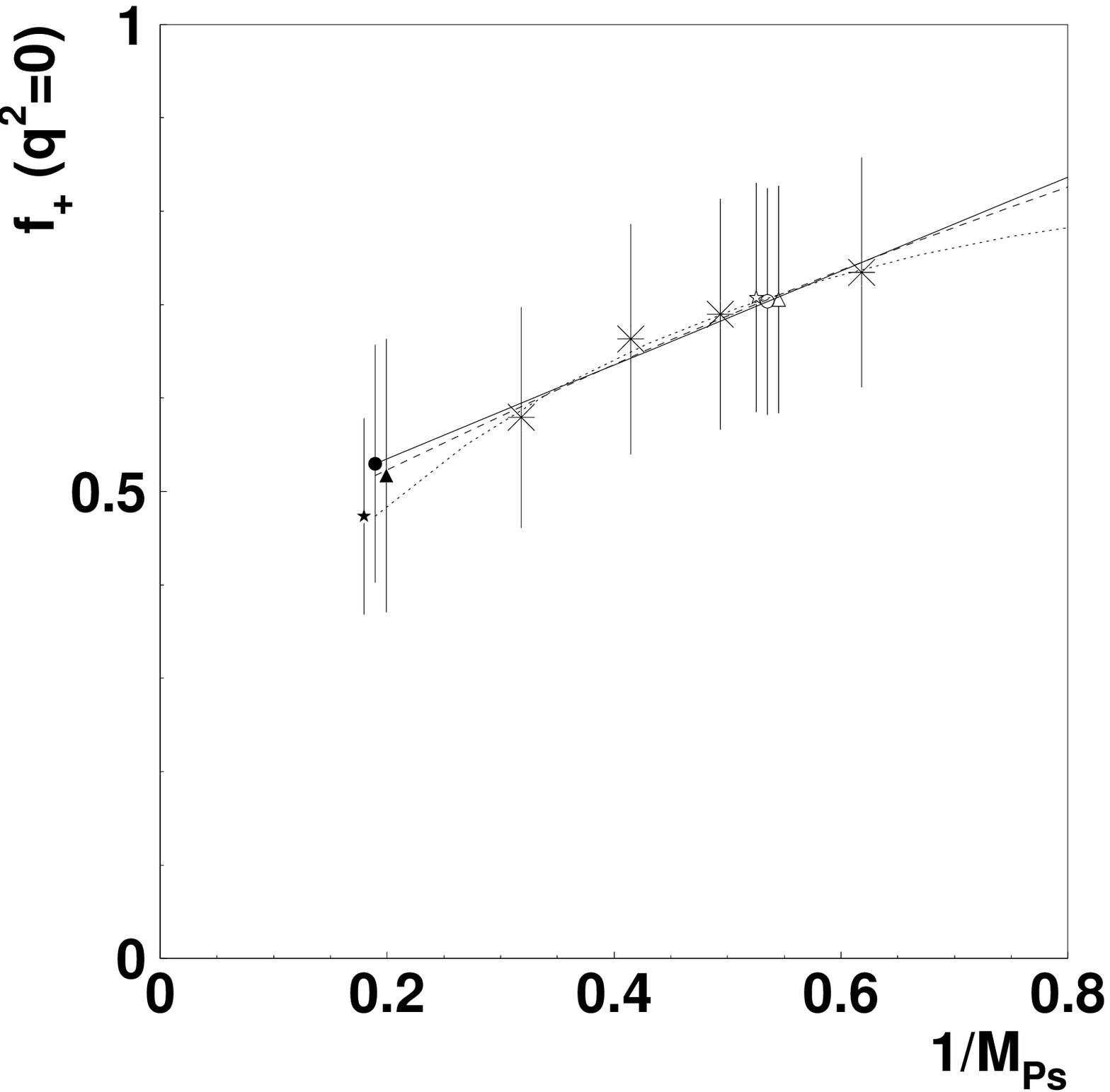}}\nolinebreak
\parbox{7.5cm}{\epsfxsize=7.5cm\epsfbox{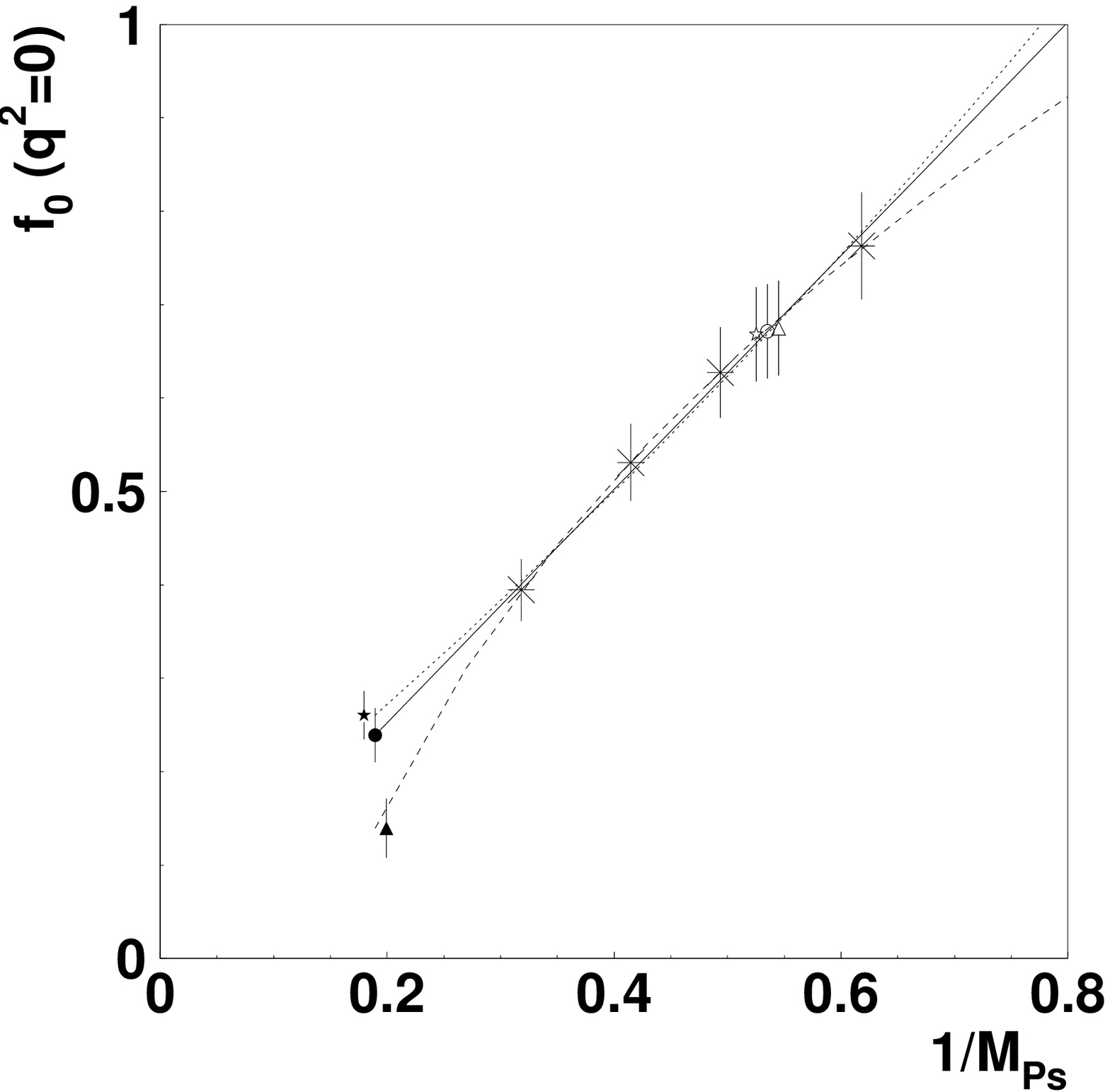}}\\
\parbox{7.5cm}{\epsfxsize=7.5cm\epsfbox{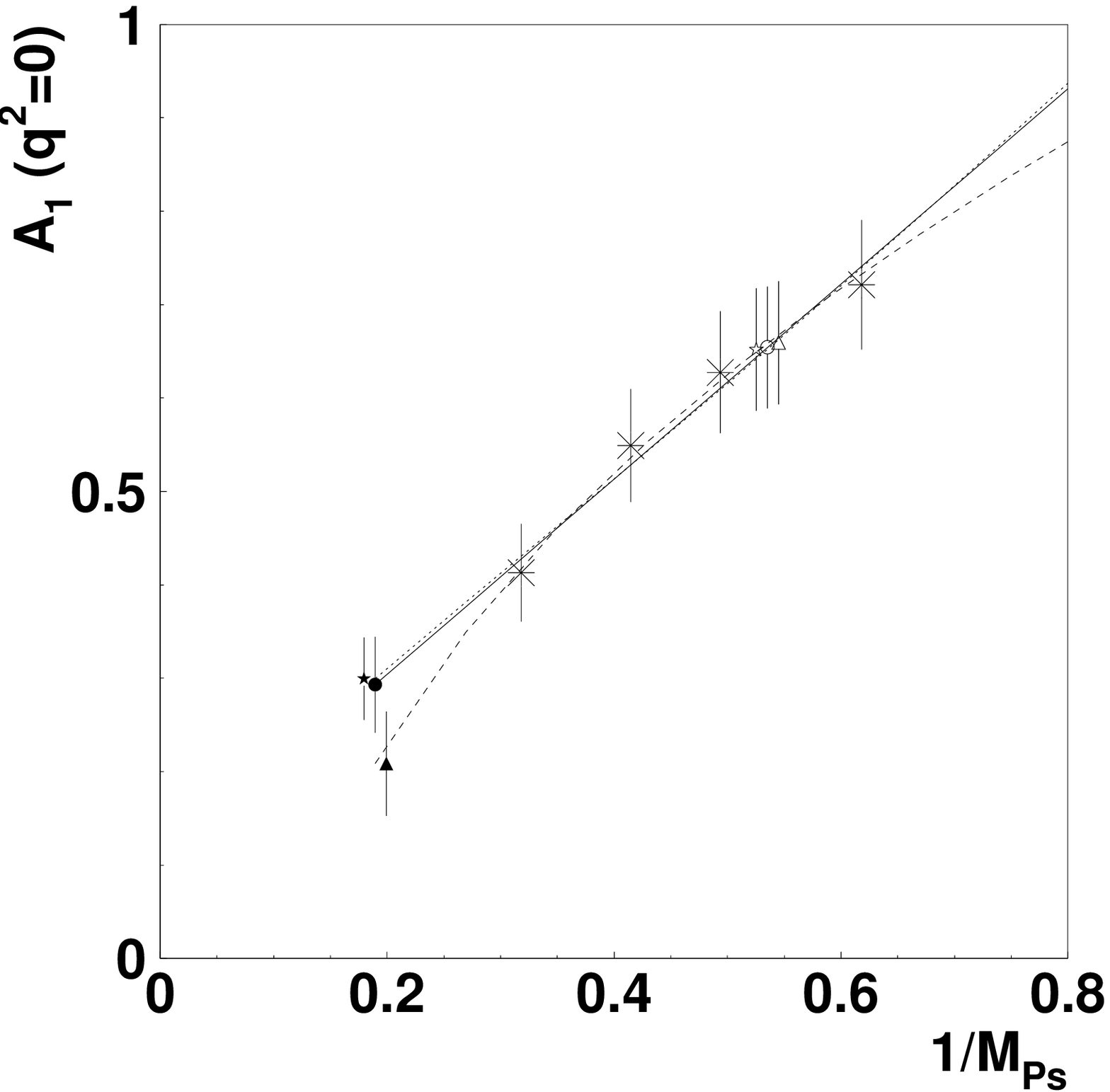}}\nolinebreak
\parbox{7.5cm}{\epsfxsize=7.5cm\epsfbox{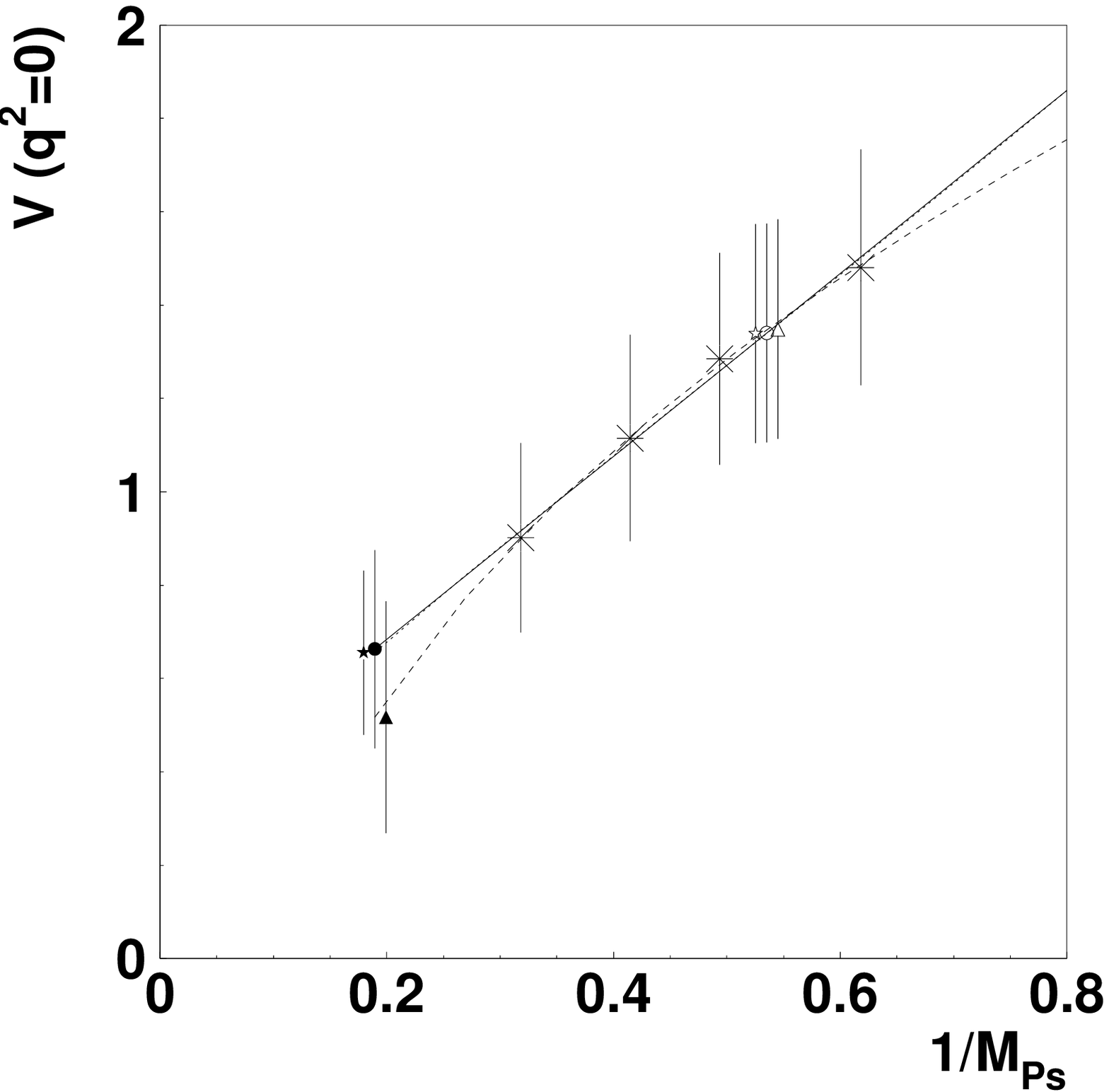}}}
\caption[a]{\label{fig_ffs_Wuppertal}
\it{Wuppertal group\cite{ffs_Wuppertal}: Mass dependence of
form factors $F(q^2=0)$ for the decays $B\rightarrow \pi$ and
$B \rightarrow \rho$. Crosses denote the
data. The solid line refers to a fit to $F = a_1 + \frac{b_1}{M_{PS}}$
(a), the dashed line to 
$F = \frac{1}{\sqrt{M_{PS}}}\left( a_2 +  \frac{b_2}{M_{PS}} \right)$
(b), and the dotted line to
$F =\sqrt{M_{PS}}\left( a_3 +  \frac{b_3}{M_{PS}} \right)$ (c).
The extrapolated results are depicted
at $1/m_{B} [\mbox{GeV}^{-1}]$ as filled circles (method a),
upper triangles (method b), and stars (method c).}}
\end{figure}
was done
considering various parameterizations for the mass dependence, as shown in
fig.\ref{fig_ffs_Wuppertal}. Unfortunately the data is still not precise
enough to discriminate between the Ans\"atze. However, the systematic 
uncertainty due to the extrapolation in mass can be estimated from
the variation of results with respect to the various Ans\"atze.

An analysis which is more guided by the Heavy Quark Effective Theory
(HQET) has been performed  by the
UKQCD\cite{ffs_light_UKQCD} collaboration. In order to extract the
maximal information from the data, they work at both, small and 
large values of $q^2$, combining the results after extrapolation
to $M_B$. Although this method assumes the validity of HQET at 
moderate meson masses, e.g. at the $D$ meson mass, it leads to quite
definite conclusions on the $q^2$ dependence of form factors 
at the $B$ meson mass.  
\begin{table}
\centering
\caption{ \it Form factors at $q^2=0$ for the semileptonic decays
$B \rightarrow \pi(l\nu)$, $B \rightarrow \rho(l\nu)$}
\vskip 0.1 in
\begin{tabular}{|l|c|c|c|c|c|} \hline
          &  $f_+(0)$ & $f_0(0)$ & $V(0)$ & $A_1(0)$ & $A_2(0)$ \\
\hline
\hline
 UKQCD\cite{ffs_light_UKQCD} 
  & $0.24^{+4}_{-3}$ 
  & $0.24^{+4}_{-3}$
  & 
  & $0.27^{+7}_{-4}\pm 3$
  & $0.28^{+9\;+4}_{-6\;-5}$            \\
 Wupp\cite{ffs_Wuppertal}
  & $0.50(14)^{+7}_{-5}$
  & $0.20(3)^{+2}_{-3}$
  & $0.61(23)^{+9}_{-6}$
  & $0.16(4)^{+22}_{-16}$       
  & $0.72(35)^{+10}_{-7}$           \\
 APE$_a$\cite{APE_ffs}
  & $0.29(6)$
  & 
  & $0.45(22)$
  & $0.29(16)$
  & $0.24(56)$                    \\
 APE$_b$\cite{APE_ffs}
  & $0.35(8)$
  & 
  & $0.53(31)$
  & $0.24(12)$
  & $0.27(80)$            \\
 ELC$_a$\cite{ELC_ffs}
  & $0.28(14)$
  & 
  & $0.37(14)$
  & $0.24(6)$
  & $0.39(24)$                    \\
 ELC$_b$\cite{ELC_ffs}
  & $0.33(17)$
  & 
  & $0.40(16)$
  & $0.21(5)$
  & $0.47(28)$            \\
\hline
\end{tabular}
\label{tab_ffs_B_to_light}
\end{table}

Table \ref{tab_ffs_B_to_light} shows the results of the different
groups\footnote{UKQCD uses $f_+(0) = f_0(0)$ as a constraint.}.
It turns out that the uncertainties are still too large to
try a reliable  extrapolation $a \rightarrow 0$.

\subsection{$B \rightarrow D,D^*$}

In order to estimate the systematic effects due to the various
inter- and extrapolation methods, the Wuppertal group has applied
the same type of analysis as for the decays $B \rightarrow \pi,\rho$.

In contrast, the UKQCD collaboration\cite{ffs_heavy_UKQCD}
has performed its analysis completely within the framework of HQET.
One major conclusion of their work is, that non-perturbative
"power corrections" to the form factors are small in the mass
region of $D$ and $B$ mesons. 
\begin{table}
\centering
\caption{ \it Branching ratios ($\%$) for exclusive semileptonic
$B$ decays with a $D$ or $D^*$ final state}
\vskip 0.1 in
\begin{tabular}{|l|c|c|} \hline
          & $\bar{B}^0 \rightarrow D^+ l^-\nu$  
          & $\bar{B} \rightarrow D^{*+} l^{-}\nu$ \\
\hline
\hline
 UKQCD
  & $1.5^{+4}_{-4}\pm 0.3$
  & $4.8^{+8}_{-9}\pm 0.5$  \\
 WUPP
  & $2.1(1.4)$
  &                          \\
 CLEO I
  & $1.8\pm 0.6\pm 0.3$
  & $4.1\pm 0.5\pm 0.7$      \\
 CLEO II
  & 
  & $4.49\pm 0.32\pm 0.39$     \\
 ARGUS
  & $2.1\pm 0.7\pm 0.6$ 
  & $4.7\pm 0.6\pm 0.6$        \\
 ALEPH
  & 
  & $5.36\pm 0.50\pm 0.76$     \\
\hline
\end{tabular}
\label{tab_branch_B_to_D}
\end{table}

In table \ref{tab_branch_B_to_D} the results for the branching
ratios are compared to the measurements\cite{Experimental_B_review}
of ARGUS, CLEO and ALEPH.

\section{Quarkonia splittings and $\alpha_s$} 
Pure heavy quark systems like quarkonia provide an ideal
laboratory to study (gluonic) inter quark forces. Lattice QCD
is well prepared to work in this lab with tools like NRQCD and
improved actions.

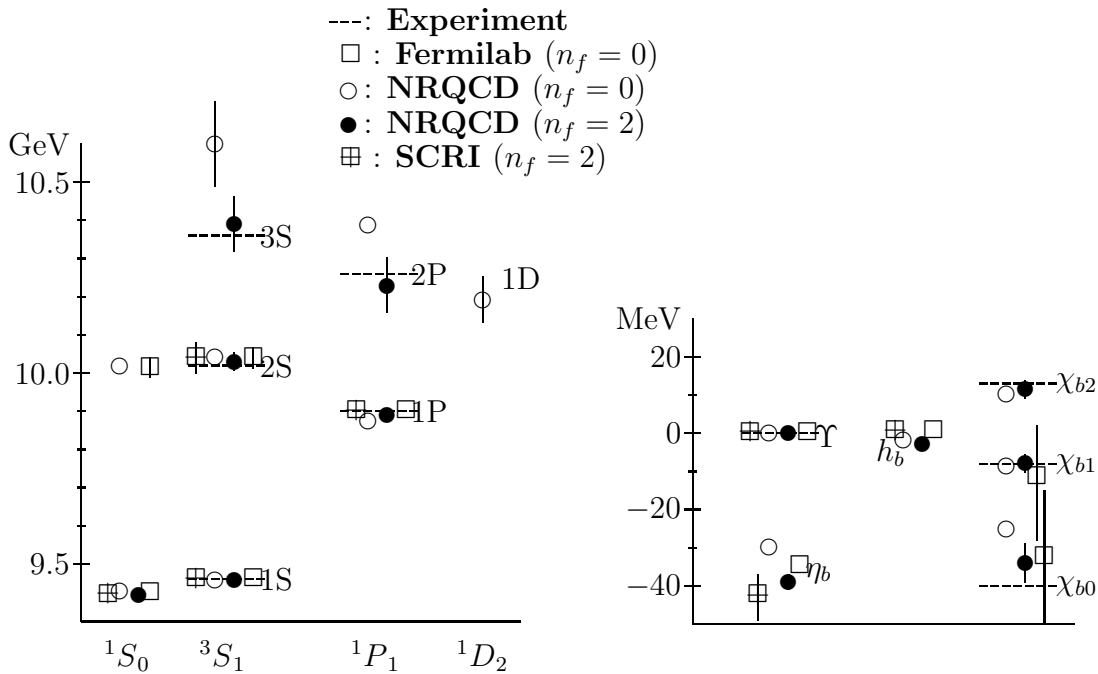
\begin{figure}[htb]
\noindent\parbox{16cm}{
\begin{minipage}[t]{7.0cm}
\setlength{\unitlength}{.02in}
\begin{picture}(130,140)(10,960)
\put(15,935){\line(0,1){125}}
\multiput(13,950)(0,50){3}{\line(1,0){4}}
\multiput(14,950)(0,10){10}{\line(1,0){2}}
\put(12,950){\makebox(0,0)[r]{9.5}}
\put(12,1000){\makebox(0,0)[r]{10.0}}
\put(12,1050){\makebox(0,0)[r]{10.5}}
\put(12,1060){\makebox(0,0)[r]{GeV}}
\put(15,935){\line(1,0){115}}


\multiput(80,1092)(3,0){3}{\line(1,0){2}}
\put(89,1092){\makebox(0,0)[l]{: {\bf Experiment}}}
\put(81,1083){\makebox(0,0)[l]{$\,\Box $ : {\bf Fermilab 
$(n_f = 0)$}}}
\put(85,1074){\makebox(0,0)[tl]{\circle{4}}}
\put(89,1074){\makebox(0,0)[l]{: {\bf NRQCD $(n_f = 0)$}}}
\put(85,1065){\makebox(0,0)[tl]{\circle*{4}}}
\put(89,1065){\makebox(0,0)[l]{: {\bf NRQCD $(n_f = 2)$}}}
\put(81,1056){\makebox(0,0)[l]{$\,\Box $ : {\bf SCRI 
$(n_f = 2)$}}}
\put(81,1056.7){\makebox(0,0)[l]{$\,+ \;\;$  }}

\put(27,930){\makebox(0,0)[t]{${^1S}_0$}}
\put(25,943.1){\circle{4}}
\put(30,942){\circle*{4}}
\put(33,942){\makebox(0,0){$\Box$}}
\put(22,941.5){\makebox(0,0){$\Box$}}
\put(22,942.5){\makebox(0,0){$+$}}

\put(25,1002){\circle{4}}
\put(33,1001){\makebox(0,0){$\Box$}}
\put(33,1001.4){\line(0,1){2.6}}
\put(33,1001.4){\line(0,-1){2.6}}

\put(52,930){\makebox(0,0)[t]{${^3S}_1$}}
\put(66,946){\makebox(0,0){1S}}
\multiput(43,946)(3,0){7}{\line(1,0){2}}
\put(50,946){\circle{4}}
\put(55,946){\circle*{4}}
\put(60,945.5){\makebox(0,0){$\Box$}}
\put(45,945.5){\makebox(0,0){$\Box$}}
\put(45,946.5){\makebox(0,0){$+$}}

\put(66,1002){\makebox(0,0){2S}}
\multiput(43,1002)(3,0){7}{\line(1,0){2}}
\put(50,1004.1){\circle{4}}
\put(55,1003){\circle*{4}}
\put(55,1004){\line(0,1){1.4}}
\put(55,1002){\line(0,-1){1.4}}
\put(60,1003.5){\makebox(0,0){$\Box$}}
\put(60,1004){\line(0,1){2.7}}
\put(60,1004){\line(0,-1){2.7}}
\put(45,1003.5){\makebox(0,0){$\Box$}}
\put(45,1004.3){\makebox(0,0){$+$}}
\put(45,1003.9){\line(0,1){4}}
\put(45,1003.9){\line(0,-1){4}}

\put(66,1036){\makebox(0,0){3S}}
\multiput(43,1036)(3,0){7}{\line(1,0){2}}
\put(50,1060){\circle{4}}
\put(50,1060){\line(0,1){11}}
\put(50,1060){\line(0,-1){11}}
\put(55,1039.1){\circle*{4}}
\put(55,1039.1){\line(0,1){7.2}}
\put(55,1039.1){\line(0,-1){7.2}}

\put(92,930){\makebox(0,0)[t]{${^1P}_1$}}

\put(106,990){\makebox(0,0){1P}}
\multiput(83,990)(3,0){7}{\line(1,0){2}}
\put(90,987.6){\circle{4}}
\put(95,989){\circle*{4}}
\put(100,989.5){\makebox(0,0){$\Box$}}
\put(87,989.5){\makebox(0,0){$\Box$}}
\put(87,990.3){\makebox(0,0){$+$}}

\put(106,1026){\makebox(0,0){2P}}
\multiput(83,1026)(3,0){7}{\line(1,0){2}}
\put(90,1038.7){\circle{4}}
\put(95,1023){\circle*{4}}
\put(95,1023){\line(0,1){7.2}}
\put(95,1023){\line(0,-1){7.2}}

\put(130,1025){\makebox(0,0){1D}}
\put(120,930){\makebox(0,0)[t]{${^1D}_2$}}
\put(120,1019.2){\circle{4}}
\put(120,1019.2){\line(0,1){6}}
\put(120,1019.2){\line(0,-1){6}}

\end{picture}
\end{minipage}

\begin{minipage}[t]{7.0cm}
\setlength{\unitlength}{.02in}
\begin{picture}(100,80)(-150,-105)

\put(15,-50){\line(0,1){80}}
\multiput(13,-40)(0,20){4}{\line(1,0){4}}
\multiput(14,-40)(0,10){7}{\line(1,0){2}}
\put(12,-40){\makebox(0,0)[r]{$-40$}}
\put(12,-20){\makebox(0,0)[r]{$-20$}}
\put(12,0){\makebox(0,0)[r]{$0$}}
\put(12,20){\makebox(0,0)[r]{$20$}}
\put(12,30){\makebox(0,0)[r]{MeV}}
\put(15,-50){\line(1,0){100}}


\multiput(28,0)(3,0){7}{\line(1,0){2}}
\put(50,2){\makebox(0,0)[t]{$\Upsilon$}}
\put(35,0){\circle{4}}
\put(40,0){\circle*{4}}
\put(45,-0.5){\makebox(0,0){$\Box$}}
\put(30,-0.5){\makebox(0,0){$\Box$}}
\put(30,0.5){\makebox(0,0){$+$}}

\put(48,-34){\makebox(0,0)[t]{$\eta_b$}}
\put(35,-29.9){\circle{4}}
\put(40,-39){\circle*{4}}
\put(40,-39){\line(0,1){2}}
\put(40,-39){\line(0,-1){2}}
\put(43,-35.3){\makebox(0,0){$\Box$}}
\put(32,-43){\makebox(0,0){$\Box$}}
\put(32,-42.3){\makebox(0,0){$+$}}
\put(32,-43){\line(0,1){6}}
\put(32,-43){\line(0,-1){6}}

\put(63,-5){\makebox(0,0)[l]{$h_b$}}
\put(70,-1.8){\circle{4}}
\put(75,-2.9){\circle*{4}}
\put(75,-2.9){\line(0,1){1.2}}
\put(75,-2.9){\line(0,-1){1.2}}
\put(78, 0.){\makebox(0,0){$\Box$}}
\put(68, 0.){\makebox(0,0){$\Box$}}
\put(68, 0.9){\makebox(0,0){$+$}}

\multiput(90,-40)(3,0){7}{\line(1,0){2}}
\put(110,-40){\makebox(0,0)[l]{$\chi_{b0}$}}
\put(97,-25.1){\circle{4}}
\put(102,-34){\circle*{4}}
\put(102,-34){\line(0,1){5}}
\put(102,-34){\line(0,-1){5}}
\put(107,-33){\makebox(0,0){$\Box$}}
\put(107,-33){\line(0,1){18}}
\put(107,-33){\line(0,-1){17}}

\multiput(90,-8)(3,0){7}{\line(1,0){2}}
\put(110,-8){\makebox(0,0)[l]{$\chi_{b1}$}}
\put(97,-8.6){\circle{4}}
\put(102,-7.9){\circle*{4}}
\put(102,-7.9){\line(0,1){2.4}}
\put(102,-7.9){\line(0,-1){2.4}}
\put(105,-12){\makebox(0,0){$\Box$}}
\put(105,-12){\line(0,1){14}}
\put(105,-12){\line(0,-1){16}}

\multiput(90,13)(3,0){7}{\line(1,0){2}}
\put(110,13){\makebox(0,0)[l]{$\chi_{b2}$}}
\put(97,10.2){\circle{4}}
\put(102,11.5){\circle*{4}}
\put(102,11.5){\line(0,1){2.4}}
\put(102,11.5){\line(0,-1){2.4}}
\end{picture}
\end{minipage}
}
\vskip -1.0cm
\caption[a]{\label{fig_Upsilon}\it $\Upsilon$ spectrum (left)
and spin splittings. The data has been compiled by
J.~Shigemitsu\cite{Upsilon}.}
\end{figure}

In order to demonstrate their  quality,
we compare in fig.\ref{fig_Upsilon}
the experimentally measured
$\Upsilon$ spectrum with the lattice results\cite{Upsilon}.
Both, the NRQCD
data and the results of the calculation with
tadpole improved clover action are in good
agreement with experiment, once dynamical fermions are included.
This sets the stage to extract the strong coupling $\alpha_s$
from the inter quark forces in the heavy quark regime.

In principle, any lattice quantity which can be reliably expanded
in powers of $\alpha$ can be used to define the strong coupling
on the lattice. In quarkonia studies one commonly\footnote{Another
ingenious choice is $\alpha_{SF}$\cite{alpha_luescher}, which is
used in connection with the Schr\"odinger functional technique. 
Unfortunately, full QCD results for $\alpha_{SF}$ are not available
yet.} works with $\alpha_P$\cite{alpha_lepage}, which is defined by
\begin{equation}
-ln W_{1,1} = \frac{4\pi}{3}\alpha_P\left(\frac{3.4}{a}\right)
[ 1 - (1.19 +0.07 n_f)\alpha_P ]\;.
\label{def_alphap}
\end{equation} 
$W_{1,1}$ is the $1 \times 1$ Wilson loop, which has to be measured on
the lattice.
In order to determine the momentum scale $Q=\frac{3.4}{a}$ of $\alpha_P$,
the lattice
cutoff $a^{-1}$ needs to be known. This is the point, where quarkonium
physics enters: $a^{-1}$ can be extracted by comparing
the measured $1S - 2S$ or $S - P$ splittings with the lattice
data. 
Finally, extrapolations in the number $n_f$ of dynamical flavours
and in the mass of dynamical quarks $m_q$ have to be carried out, 
$n_f=0,2 \rightarrow n_f=3$, 
$m_q \rightarrow m_{eff} = (m_u+m_d+m_s)/3$.
As a result of this procedure, the NRQCD collaboration 
quotes\cite{alphap_NRQCD}
$$
\alpha_P^{(n_f=3)}[8.2\mbox{GeV}] = \left\{ \begin{array}{ll}
0.1948(29)(11)(37) & \mbox{S - P splitting} \\
\\
0.1962(41)(08)(40) & \mbox{1S - 2S splitting} 
\end{array}
\right. \; ,
$$
where the first error accounts for statistical uncertainties,
the second for discretization effects and the third for 
uncertainties due to the extrapolations.
To compare with other determinations of the strong coupling it is
advantageous to convert $\alpha_P$ to $\alpha_{\overline{MS}}$.
Unfortunately, the corresponding two loop conversion formula has
been calculated\cite{luescher_2loop} hitherto  only for $n_f=0$.
This leads to an additional systematic uncertainty, which has to
be taken into account.
At the $M_{Z0}$ mass, the NRQCD collaboration finds\cite{Upsilon}
$$
\alpha_{\overline{MS}}[M_{Z0}] = \left\{ \begin{array}{ll}
0.1175(11)(13)(19) & \mbox{S - P splitting} \\
\\
0.1180(14)(14)(19) & \mbox{1S - 2S splitting} 
\end{array}
\right. \; .
$$ 
Note that the last error, which accounts for the uncertainty in
the conversion, dominates.

\section{Summary and Conclusions}

We have discussed the importance of recent developments and results 
in heavy quark physics on the lattice. Considerable progress 
has been achieved with respect to the reliability of
the continuum extrapolation and to the inclusion of
internal fermion loops. This is mainly
due to the dramatic improvement in computer performance.

 In order to reduce discretization effects, a variety of promising new
ideas and methods has been developed. With the  advent of fast
parallel computers, the benefit of these methods can be tested reliably
in numerical simulations. Once this has been done, they will
be used for precise calculations at or even beyond the $b$ quark
mass.  

\section{Acknowledgements}

I thank Apoorva Patel for useful discussions. 
\end{document}